\documentclass[ 
aps,%
10pt,%
final,%
notitlepage,%
oneside,%
onecolumn,%
nobibnotes,%
nofootinbib,%
superscriptaddress,%
noshowpacs,%
centertags]%
{revtex4-2}

\usepackage{enumitem}

\usepackage{tikz}
\usepackage[mathscr]{eucal}
\usepackage{hyperref}
\usepackage{amsmath}
\usepackage{amssymb}
\usepackage{cancel}
\usepackage{mathtools}
\usepackage{physics}

\begin{document}
	
\title{Cosmological Constant Suppression in Non-Stationary Scalar Covariant State}
	
\begin{abstract}
We study a spatial-temporal structure of quantum fluctuations in the stress-energy tensor of zero-point modes for a scalar field in order to formulate a covariant model. The model describes an invariant vacuum contribution to the cosmological constant in the non-stationary coherent state in a finite volume. Bare and effective mean values of vacuum energy density are compared. 
\end{abstract}

	\author{\firstname{Asya}~\surname{Aynbund}}
	\email{aynbund.asya@phystech.edu}
	\affiliation{Moscow Institute of Physics and Technology,
		Institutsky lane 9, Dolgoprudny, Moscow region, 141700, Russia}
	
	\author{\firstname{V.V.}~\surname{Kiselev}}
	\email{kiselev.vv@phystech.edu; Valery.Kiselev@ihep.ru}
	
	\affiliation{Moscow Institute of Physics and Technology,
    Institutsky lane 9, Dolgoprudny, Moscow region, 141700, Russia}
	\affiliation{State Research Center of the Russian Federation  ``Institute for High Energy
	Physics'' of National Research Centre  ``Kurchatov Institute'',
    Nauki 1, Protvino, Moscow Region, 142281, Russia}	
	\maketitle
\section{Introduction}	
Despite recent experimental indications provided by Dark Energy Spectroscopic Instrument (DESI) on the baryonic acoustic oscillations signalling for the significant dependence of dark energy equation of state on the scale factor gouvering the Universe expansion \cite{DESI:2024uvr,DESI:2024lzq,DESI:2024mwx}, the concordance model $\Lambda$CDM as based on the cosmological constant and cold dark matter, still composes the ground to consider the physical meaning of dimensional parameters determining the observed reality. In this respect one could speculate on some dark energy dynamics as extra small but currently visible corrections to the main effect given by the cosmological constant. Then the nature of energy scale setting cosmological constant continues to be of fundamental interest \cite{Weinberg:1988cp,Zeldovich:1968ehl}.

It has been experimentally established that the energy scale, which sets the cosmological constant or vacuum energy density, is $10^{-3}$ eV according to cosmological data on the evolution of the Universe \cite{Baumann:2018muz}. The problem is that such a value considers to be unnatural or even occasional in theoretical models including the landscape in the string theory \cite{Halverson:2018vbo}, where the natural scale would be of the order of $10^{18}$ GeV \cite{Kinney:2009vz}. In the present paper we propose a model describing an origin for the suppression of cosmological constant scale. The mechanism of suppression is the following: a scalar field in a finite volume in a non-stationary coherent state possesses an isotropic stress-energy tensor, while the contribution of its vacuum state with zeroth number of quanta $\ket{0}$ is suppressed exponentially in the coherent state if an average number of quanta is much greater than one \cite{Balitsky:2016gwm,Balitsky:2014epa,Kiselev:2013zia}. For the equation of state parameter being the ratio of pressure to the energy density $w={p}/{\rho}=-1$, it is necessary to consider 4D-isotropic quantum fluctuations.

The paper structure is the following: in Section~\ref{sec:Fluct} zero-point modes of scalar field are described and the necessity of spatial-temporal covariancy is established. Section~\ref{sec:CutOff} deals with the case of scalar field in a finite volume. We argue for a covariant model and non-stationary coherent state to derive our main observation. We stress a consistency of scalar field with the inflaton in the early Universe to involve natural values of input parameters in the dynamics. Section~\ref{sec:NumEst} is devoted to a numerical estimates relevant to the problem of cosmological constant scale. In Conclusion we summarize the results.

\section{Coherent spatial-temporal structure of fluctuations} \label{sec:Fluct}
\subsection{Zero-point modes of scalar field and spatial covariance}
Under the secondary quantisation of scalar field one can write the expression for the stress-energy tensor averaged over the zero-point modes, i.e. in the stationary state $|0\rangle$ in which there are no quanta of the field,
\begin{equation}
\label{zp-1}
	\langle 0 |T_{\mu\nu}|0\rangle =\int\frac{\mathrm{d}^3k}{(2\pi)^32k_0}\,
	k_\mu k_\nu,\quad\mbox{at } \mu\mbox{ and }\nu \in \big\{\overline{0,3}\big\},
\end{equation}
where in we have defined the relativistically invariant normalisation of creation and annihilation operators of field quanta with a definite momentum $\boldsymbol k$ by the following commutation relation:
\begin{equation}
\label{zp-2}
	[\hat a(\boldsymbol k),\hat a(^\dagger\boldsymbol k')]=(2\pi)^3
	\delta(\boldsymbol k-\boldsymbol k')\cdot 2k_0,\qquad k_0=\sqrt{\boldsymbol k^2+m^2}.
\end{equation}
In order to get final values in (\ref{zp-1}) one introduce a cut-off of momenta, that we put to the sub-planckean scale $\Lambda_P$. Moreover, we postulate the spatial covariance, hence, $\mbox{at } \alpha\mbox{ and }\beta \in \big\{\overline{1,3}\big\}$ 
\begin{equation}
\label{zp-3}
	\langle 0 |T_{\alpha\beta}|0\rangle =
	\int\limits_{V(\Lambda_P)}\frac{\mathrm{d}^3k}{(2\pi)^32k_0}\,
	k_\alpha k_\beta=\int\limits_{V(\Lambda_P)}\frac{\mathrm{d}^3k}{(2\pi)^32k_0}\,
	\frac13\,\delta_{\alpha\beta}\boldsymbol k^2,
\end{equation}
while
\begin{equation}
\label{zp-4}
	\langle 0 |T_{0\beta}|0\rangle =
	\int\limits_{V(\Lambda_P)}\frac{\mathrm{d}^3k}{(2\pi)^32k_0}\,
	k_0 k_\beta=0,
\end{equation}
and 
\begin{equation}
\label{}
\label{zp-4_new}
	\langle 0 |T_{00}|0\rangle =
	\int\limits_{V(\Lambda_P)}\frac{\mathrm{d}^3k}{(2\pi)^32k_0}\,
	k_0 k_0= \int\limits_{V(\Lambda_P)}\frac{\mathrm{d}^3k}{2(2\pi)^3}\,
	\sqrt{k^2+m^2}.
\end{equation}
For an invariant vacuum $\langle 0 |T_{\alpha\beta}|0\rangle = -\delta _{\alpha\beta} \langle 0 |T_{00}|0\rangle,$ that is  not possible for \ref{zp-3} and \ref{zp-4_new}.
So, one can calculate integrals in the explicit forms in some cases.

\textit{Relativistic limit.} For instance, at $\Lambda_P\gg m$ under the expansion
$$
	k_0\approx k+\frac{m^2}{2k}+\ldots
$$
at $k=|\boldsymbol k|$ one gets the ultra-relativistic expressions 
\begin{equation}
\label{zp-5}
	\langle 0 |T_{\alpha\beta}|0\rangle \approx \frac{1}{3}\,\delta_{\alpha\beta}
	\int\limits_0^{\Lambda_P}\frac{4\pi k^2\mathrm{d}k}{(2\pi)^3}\,
	\frac{k^2}{2k}\Big(1-\frac{m^2}{2k^2}\Big)= \frac{1}{3}\,\delta_{\alpha\beta}\cdot
	\frac{\Lambda_P^4}{16\pi^2}\Big(1-\frac{m^2}{\Lambda_P^2}\Big),
\end{equation}
while
\begin{equation}
\label{zp-6}
	\langle 0 |T_{00}|0\rangle \approx 
	\int\limits_0^{\Lambda_P}\frac{4\pi k^2\mathrm{d}k}{(2\pi)^3}\,
	\frac{k}{2}\Big(1+\frac{m^2}{2k^2}\Big)=
	\frac{\Lambda_P^4}{16\pi^2}\Big(1+\frac{m^2}{\Lambda_P^2}\Big).
\end{equation}

Therefore, the ratio of isotropic pressure $p=p_x= p_y= p_z=
T_{xx}=T_{yy}=T_{zz}$ to the energy density $\rho=T_{00}$ determines the state parameter
\begin{equation}
\label{zp-w}
	w=\frac{p}{\rho}\approx\frac13\Big(1-\frac{2 m^2}{\Lambda_P^2}\Big)\to 
	w_\mathrm{rad}=\frac13\quad\mbox{at }\frac{m\hspace*{2mm}}{\Lambda_P}\to 0.
\end{equation}

\textit{Non-relativistic limit.} Analogously at $\Lambda_P\ll m$ under the expansion
$$
	k_0\approx m+\frac{k^2}{2m}+\ldots
$$
at $k=|\boldsymbol k|$ one gets the non-relativistic expressions 
\begin{equation}
\label{zp-5n}
	\langle 0 |T_{\alpha\beta}|0\rangle \approx \frac{1}{3}\,\delta_{\alpha\beta}
	\int\limits_0^{\Lambda_P}\frac{4\pi k^2\mathrm{d}k}{(2\pi)^3}\,
	\frac{k^2}{2m}\Big(1-\frac{k^2}{2m^2}\Big)= \frac{1}{3}\,\delta_{\alpha\beta}\cdot
	\frac{m\Lambda_P^3}{4\pi^2}\Big(\frac{\Lambda_P^2}{5 m^2}-
	\frac{\Lambda_P^4}{14 m^4}\Big),
\end{equation}
while
\begin{equation}
\label{zp-6n}
	\langle 0 |T_{00}|0\rangle \approx
	\int\limits_0^{\Lambda_P}\frac{4\pi k^2\mathrm{d}k}{(2\pi)^3}\,
	\frac{m}{2}\Big(1+\frac{k^2}{2m^2}\Big)=
	\frac{m\Lambda_P^3}{4\pi^2}\Big(\frac13+\frac{\Lambda_P^2}{10 m^2}\Big).
\end{equation}

Therefore, the ratio of isotropic pressure $ p$ to the energy density $\rho$ determines the state parameter in the non-relativistic limit,
\begin{equation}
\label{zp-wn}
	w=\frac{ p}{\rho}\approx \frac35\, \frac{\Lambda_P^2}{m^2}
	\Big(1-\frac{23}{35}\,\frac{\Lambda_P^2}{m^2}\Big)\to 
	w_\mathrm{m}=0\quad\mbox{at }\frac{\Lambda_P}{m\hspace*{2mm}}\to 0.
\end{equation}

Both limits of radiation and dust matter in (\ref{zp-w}) and (\ref{zp-wn}) correspondingly show that the spatially covariant expressions for zero-point modes do not give stress-energy tensors fitting the invariant vacuum with $ T_{\mu\nu}(\mathrm{vac})=\rho_\mathrm{vac} g_{\mu\nu}$. On the contrary, zero-point fluctuations give the stress-energy tensors of particles as depends on the value of cut-off. This means that the state $|0\rangle$ under such fluctuation model is not the vacuum. Therefore, one needs to change the fluctuation model to get the stress-energy tensor of vacuum.\\[-5mm]
\begin{figure}[h]
\begin{center}
\includegraphics[width=0.9in]{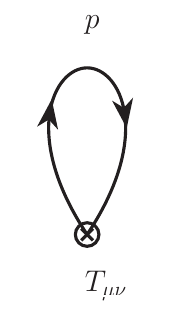}\\[-3mm]
\caption{Feynman diagram for averaging the stress-energy operator over the vacuum.}
\label{loop-T}
\end{center}
\end{figure}\\[-9mm]

The known way is to use the Feynman propagator in four-dimensional notations 
\begin{equation}
\label{prop-1}
	\langle 0|\hat{\mathcal  T}\varphi(x)\varphi(y)|0\rangle= 
	\int \frac{\mathrm{d}^4k}{(2\pi)^4}\,\frac{\mathrm{i}}{k^2-m^2+\mathrm{i}0}
	\,\mathrm{e}^{-\mathrm{i}(x-y)k}
\end{equation}
 at $\hat{\mathcal{T}}\varphi(x)\varphi(x)=\varphi(x)\varphi(x)$ in the the averaging the stress-energy operator 
 \begin{equation}
\label{prop-2}
	\bra {0} T_{\mu\nu}(x) \ket {0}   =\langle 0| \big(\partial_\mu\varphi(x)\big)
	\big(\partial_\nu\varphi(x)\big)-\frac12 g_{\mu\nu}\big(\big(\partial_\lambda\varphi(x)\big)^2-
	m^2\varphi^2(x)\big)|0\rangle.
\end{equation}
Then following Feynman rules for diagram shown on Fig. \ref{loop-T} one gets 
\begin{equation}
\label{prop-3}
	\langle 0 |T_{\mu\nu}(x)|0\rangle =
	\int \frac{\mathrm{d}^4k}{(2\pi)^4}\,\frac{\mathrm{i}}{k^2-m^2+\mathrm{i}0}
	\Big(k_\mu k_\nu-\frac12\,g_{\mu\nu}(k^2-m^2)\Big)
	\mathrm{e}^{-\mathrm{i}(x-y)k}\Big|_{y\to x}.
\end{equation}
The limit of $y\to x$ implies that a single pole in the propagator contributes and by the formula with the principal value and delta-function
\begin{equation}
\label{prop-4}
	\,\frac{\mathrm{i}}{k^2-m^2+\mathrm{i}0}=
	\mathtt{PV}\,\frac{\mathrm{i}}{k^2-m^2}+\pi\,
	\delta\big(k^2-m^2\big)
\end{equation}
one arrives to the same expression obtained by the secondary quantisation as presented in (\ref{zp-1}) since the purely imaginary complex contribution should be equal to zero.

During the mechanism of working with the Feynman diagrams the Wick rotation leads to time and spatial coordinates 
being equal in rights. Nevertheless, we can suppose that a true spatial-temporal structure of quantum fluctuations follows an other covariant form as provided by the Wick rotation in the plain of temporal component of momentum.

\subsection{Scalar field and spatial-temporal covariancy}
Let us set $k_0=\mathrm{i}k_4$ by the Wick rotation that implements the four dimensional Euclidean space structure with the metric $g^E_{\mu\nu}=\delta_{\mu\nu}$ due to $g_{\mu\nu}\mapsto -g^E_{\mu\nu}$. Then $k^2=-k_4^2-\boldsymbol k^2=-k_E^2$ and 
\begin{equation}
\label{prop-5}
	\langle 0 |T_{\mu\nu}(x)|0\rangle \mapsto 
	\langle \mbox{vac} |T_{\mu\nu}(x)|\mbox{vac}\rangle = 
	\int \frac{\mathrm{d}^4k_E}{(2\pi)^4}\,\frac{1}{k_E^2+m^2}
	\Big(k_\mu^E k_\nu^E-\frac12\,g^E_{\mu\nu}(k_E^2+m^2)\Big)
\end{equation}
with no poles. In this way the four-dimensional covariance implies that in the integranl of (\ref{prop-5}) one can apply the following replacement:
\begin{equation}
\label{cov-1}
	k_\mu^E k_\nu^E\mapsto \frac14\,g^E_{\mu\nu}k_E^2.
\end{equation}
Therefore, 
\begin{equation}
\label{prop-6}
	\langle \mbox{vac} |T_{\mu\nu}(x)|\mbox{vac}\rangle  = -\frac14\,g^E_{\mu\nu}
	\int \frac{\mathrm{d}^4k_E}{(2\pi)^4}\left(1+\frac{m^2}{k_E^2+m^2}\right)=
	 \frac14\,g_{\mu\nu}
	\int \frac{\mathrm{d}^4k_E}{(2\pi)^4}\left(1+\frac{m^2}{k_E^2+m^2}\right).
\end{equation}
Hence, considering components of \ref{prop-6}:
\begin{equation}
\label{prop-6_new}
	T_{\alpha\beta} = -\delta_{\alpha\beta}T_{00} = p \delta_{\alpha\beta}, \quad T_{00} = \rho
\end{equation}
This expression explicitly satisfies the vacuum equation of state
\begin{equation}
\label{vac-1}
	w_\mathrm{vac}=-1.
\end{equation}
Introducing the cut-off $\Lambda_E$ one gets a finite positive value of vacuum energy density
\begin{equation}
\label{vac-2}
	\rho_\mathrm{vac}=\langle \mbox{vac} |T_{00}|\mbox{vac}\rangle =
	\frac14\int\frac{\mathrm{d}^3\Omega_E}{(2\pi)^4}\int\limits_0^{\Lambda_E}
	k_E^3\mathrm{d}k_E
	\left(1+\frac{m^2}{k_E^2+m^2}\right)=\frac{\Lambda_E^4}{128\pi^2}
	\left(1+\frac{2m^2}{\Lambda_E^2}-\frac{2m^4}{\Lambda_E^4}
	\ln\left(1+\frac{\Lambda_E^2}{m^2}\right)\right).
\end{equation}

Thus, the true coherent structure of spatial-temporal quantum fluctuations in the vacuum supposes the isotropic 4-dimensional formulation.

\section{Coherent spatial-temporal structure in finite volume with scale cut-off}\label{sec:CutOff}
\subsection{Finite volume and spatial structure} 
Consider the scalar field depending on time in a finite volume. The action
\begin{equation}
\label{fin-1}
	S=\int\limits_{V_{[3]}}\mathrm{d}^3r\int\mathrm{d}t\, \frac12\,
	\big((\partial_\mu\varphi(x))^2-m^2\varphi^2(x)\big)
\end{equation}
at $\varphi(x)\mapsto \varphi(t)$ is reduced to the action of harmonic oscillator
\begin{equation}
\label{fin-2}
	S=V_{[3]}\int\mathrm{d}t\, \frac12\,
	\big((\partial_t\varphi(t))^2-m^2\varphi^2(t)\big)
\end{equation}
comparing with the action for the oscillator:
\begin{equation}
\label{act_osc}
	S_\mathrm{osc}=\int\mathrm{d}t\, \left(\frac{M_\mathrm{osc}\dot q_\mathrm{osc}^2}{2} - \frac{M_\mathrm{osc}\omega^2_\mathrm{osc}q_\mathrm{osc}^2}{2} \right)
\end{equation}
and using considerations of dimensionality of quantities, namely $[q_\mathrm{osc}] = \frac{1}{[E]}, [\varphi] = [E], \Lambda = [E],$ hence the oscillator coordinate, mass and frequency equal to 
\begin{equation}
\label{fin-3}
	q_\mathrm{osc}=\frac{\varphi}{\Lambda^2},\qquad M_\mathrm{osc} = \Lambda^4 V_{[3]},
	\qquad \omega_\mathrm{osc}=m,
\end{equation}
where we have introduce a scale of energy $\Lambda$. Since under the standard quantisation in the stationary state with no quanta
\begin{equation}
\label{fin+4}
	\langle 0 | q_\mathrm{osc}^2|0\rangle=\frac{1}{2M_\mathrm{osc}\omega_\mathrm{osc}},
	\qquad \langle 0| M^2_\mathrm{osc}(\partial_t q_\mathrm{osc})^2|0\rangle=\frac12\,
	M_\mathrm{osc}\omega_\mathrm{osc},
\end{equation}
using the virial theorem for the oscillator we get
\begin{equation}
\label{fin-4}
	\langle 0| m^2\varphi^2|0\rangle = \langle 0|(\partial_t\varphi)^2|0\rangle = 
	\frac12\,\frac{m}{V_{[3]}},
\end{equation}
that allows us to calculate the bare values of stress-energy tensor components in the stationary state. So, the energy density (also due to the use of the virial theorem) equals
\begin{equation}
\label{fin-5}
	\langle 0| T_{00}|0\rangle =\frac12\langle 0| (\partial_t\varphi)^2+ m^2\varphi^2 |0\rangle =
	\frac12\,\frac{m}{V_{[3]}},
\end{equation}
while the pressure in the isotropic case is given by 
\begin{equation}
\label{fin-6}
	T_{xx} = T_{yy} = T_{zz} = p=\frac12\langle 0| (\partial_t\varphi)^2- m^2\varphi^2 |0\rangle = 0.
\end{equation}
Thus, the stationary state with no quanta and separately described time and finite volume does not correspond to the vacuum because the parameter in the equation of state fits the dust, i.e. particles with zeroth pressure.

\subsection{Covariant model}
Let us set the field depending on the proper time $\tau$ in the finite volume in the covariant four-dimensional form, so that the action equals
\begin{equation}
\label{tau-1}
	S=\int\limits_{V_4,\Delta\tau}\mathrm{d}^4 x\,\frac12
	\big((\partial_\tau\varphi)^2\partial_\mu\tau\partial_\nu\tau\,g^{\mu\nu}-m^2\varphi^2\big)
	\xrightarrow[\Lambda \text{ and } \langle \cdot \rangle]{\text{cut-off}} S_\Lambda=
	\frac{1}{\Lambda^3}\int\mathrm{d}\tau \frac12
	\Big((\partial_\tau\varphi)^2\big\langle \partial_\mu\tau\partial_\nu\tau\big\rangle 
	g^{\mu\nu}-m^2\varphi^2\Big),
\end{equation}
where we have introduced the transition to the invariant cut-off $\Lambda$ determining the finite volume factor in the action, while the covariant spatial-temporal structure in the model is defined by the following expression:
\begin{equation}
\label{tau-2}
	\big\langle \partial_\mu\tau\partial_\nu\tau\big\rangle = \frac{1}{4}g_{\mu \nu}
\end{equation}
a complete analogy with the isotropy condition in relation (\ref{cov-1}) under the following normalization condition setting the scale of invariant variable $\tau$:
\begin{equation}
\label{tau-2a}
	\langle(\partial_\lambda\tau)^2\rangle=1,
\end{equation}
in a complete analogy with (\ref{cov-1}).  Then the effective action\footnote{Another sort of argumentation based on the fact of effective covariant cut-off in the momentum space for the field fluctuations, will be presented elsewhere.} equals
\begin{equation}
\label{cov-2}
	S_\Lambda=\frac{1}{\Lambda^3}\int\mathrm{d}\tau \frac{1}{2}\left((\partial_\tau \varphi)^2 -m^2\varphi^2\right)
\end{equation}
which is the action of harmonic oscillator, when
\begin{equation}
\label{cov-3}
	q_\mathrm{osc}=\frac{\varphi}{\Lambda^2},\qquad M_\mathrm{osc} = \Lambda,
	\qquad \omega_\mathrm{osc}=m.
\end{equation}

In this way the stress-energy tensor is transformed to 
\begin{multline}
\label{cov-4}
	T_{\mu\nu}=\big(\partial_\mu\varphi(x)\big)
	\big(\partial_\nu\varphi(x)\big)-\frac12 g_{\mu\nu}\big(\big(\partial_\lambda\varphi(x)\big)^2-
	m^2\varphi^2(x)\big) = \\[2mm] {}  =
	(\partial_\mu\tau)
	(\partial_\nu\tau)(\partial_\tau\varphi)^2-\frac12 g_{\mu\nu} 
	\big((\partial_\lambda\tau)^2(\partial_\tau\varphi)^2- 
	m^2\varphi^2\big) \mapsto \\[2mm] 
	{} \mapsto  	\quad
	\langle \mbox{vac} |T_{\mu\nu}^\mathrm{\Lambda}|\mbox{vac}\rangle
	=\langle \mbox{vac} |
	\big\langle \partial_\mu\tau\partial_\nu\tau\big\rangle (\partial_\tau\varphi)^2-
	\frac12 g_{\mu\nu} \big(\langle(\partial_\lambda\tau)^2\rangle
	(\partial_\tau\varphi)^2-
	m^2\varphi^2\big)|\mbox{vac}\rangle,
\end{multline}
hence, under (\ref{tau-2}) and (\ref{tau-2a}) with the evident  application of (\ref{fin+4}) we finally get
\begin{equation}
\label{cov-5}
	\langle \mbox{vac} |T_{\mu\nu}^\mathrm{\Lambda}|\mbox{vac}\rangle=g_{\mu\nu}\langle \mbox{vac} |\big(-\frac14\,(\partial_\tau\varphi)^2+\frac12 m^2\varphi^2\big)|\mbox{vac}\rangle=g_{\mu\nu} \frac18\,m\Lambda^3. 
\end{equation}
Thus, the covariant structure of space-time in the finite volume results in the vacuum-like stress-energy tensor at least for the basic state with no quanta of oscillator and the bare value of energy density
\begin{equation}
\label{cov-6}
	\rho_\mathrm{vac}^\mathrm{bare}=\langle \mbox{vac} |T_{00}^\Lambda |\mbox{vac}\rangle =\frac18\,m\Lambda^3>0.
\end{equation}

\subsection{Non-stationary coherent state}
Since the universe evolves, it is natural to put the scalar field in the finite volume to be in a non-stationary state. Moreover, if the state appears as a result of eventual fluctuation with respect to vacuum state, then the number of quanta are distributed by Poisson, while the state is coherent. Therefore, the contribution of vacuum state to the average stress-energy tensor is suppressed as
\begin{equation}
\label{cov-7}
	\langle T_{\mu\nu}^\Lambda\rangle_\mathrm{vac}= 
	\langle \mbox{vac} |T_{\mu\nu}^\Lambda	|\mbox{vac}\rangle \cdot
	\mathrm{e}^{-\langle n\rangle}=\rho_\mathrm{vac}^\mathrm{bare}\cdot
	\mathrm{e}^{-\langle n\rangle}=\frac18\,m\Lambda^3\cdot\mathrm{e}^{-\langle n\rangle},
\end{equation}
where $\langle n \rangle$ is an average number of quanta in the coherent state. 

In this way the energy of coherent state in the final volume in accordance with action (\ref{cov-2}) is given by 
\begin{equation}
\label{cov-8}
	\langle E\rangle= \omega_\mathrm{osc}\left(\langle n\rangle+\frac12\right)=
	m \left(\langle n\rangle+\frac12\right),
\end{equation}
while the same quantity in terms of energy density within a finite volume $V_{[3]}$ is equal to 
\begin{equation}
\label{cov-9}
	\langle E\rangle= V_{[3]}\,\rho_\mathrm{vac}^\mathrm{bare}
	\left(\langle n\rangle+\frac12\right)\cdot 2
	=V_{[3]}\,\frac14\,m \Lambda^3 \left(\langle n\rangle+\frac12\right),
\end{equation}
since
$$
	\frac12\,m=V_{[3]}\,\rho_\mathrm{vac}^\mathrm{bare}.
$$
Therefore, the reference volume equals
$$
	V_{[3]}=\frac{4}{\Lambda^3}.
$$

\subsection{Interaction with gravity} 
We have considered the model of free scalar field in finite volume with no gravity. Introduce the following term of non-minimal interaction
\begin{equation}
\label{int-1}
	S_\mathrm{int}=-\frac12 \,M_\mathrm{int}\int\mathrm{d}^4x\,\sqrt{-g}\,R\cdot\varphi=
	-\frac12\,\tilde m_\mathrm{Pl}^2\int\mathrm{d}^4x\,\sqrt{-g}\,R\cdot
	\frac{\varphi\hspace*{2mm}}{\Lambda_\mathrm{int}},
\end{equation}
where the additional scale $M_\mathrm{int}$ is written in the form corresponding 
to the Einstein--Hilbert action
\begin{equation}
\label{int-2}
	S_\mathrm{EH}=-\frac12\,\tilde m_\mathrm{Pl}^2\int\mathrm{d}^4x\,\sqrt{-g}\,R
\end{equation}
with the reduced Planckean mass $\tilde m_\mathrm{Pl}$ related with the Newton constant $G$,
\begin{equation}
\label{int-3}
	\tilde m_\mathrm{Pl}=(8\pi G)^{-1}.
\end{equation}
So, the sum of these actions equals
$$
	S_\mathrm{int}+S_\mathrm{EH}=
	-\frac12\,\tilde m_\mathrm{Pl}^2\int\mathrm{d}^4x\,\sqrt{-g}\,R\cdot
	\Big(1+\frac{\varphi\hspace*{2mm}}{\Lambda_\mathrm{int}}\Big),
$$
and $\Lambda_\mathrm{int}$ determines the measure of deviation from the Einstein--Hibert action in terms of ratio of scalar field value to the scale  $\Lambda_\mathrm{int}$. It is natural to put this scale to be of the same order of magnitude to $\Lambda_\mathrm{int}$. Due to the non-minimal interaction the sum $S_\mathrm{int}+S_\mathrm{EH}$ represents the action in the Jordan frame. One can introduce the conformal transformation with the following conformal factor $\Omega(\varphi)$ depending on the field:
\begin{equation}
\label{int-4}
	\Omega(\varphi)=1+\frac{\varphi\hspace*{2mm}}{\Lambda_\mathrm{int}}
\end{equation}
determining the transformation of metric by 
\begin{equation}
\label{int-5}
	g_{\mu\nu}\mapsto\frac{1}{\Omega(\varphi)}\, \mathrm g_{\mu\nu}.
\end{equation}
Then the action of scalar field and new metrics gets the following form in the Einstein frame \cite{Kallosh:2013tua}:
\begin{equation}
\label{int-6}
	S=\int\mathrm{d}^4x\,\sqrt{-\mathrm g}\,
	\Big(-\frac12\,\tilde m_\mathrm{Pl}^2 \mathrm R+\frac12(\partial_\mu\varphi)^2
	\left(\frac{1}{\Omega(\varphi)}+\frac32\,\tilde m_\mathrm{Pl}^2
	\Bigg(\frac{\partial \ln\Omega}{\partial\varphi}\Bigg)^2
	\right)-V_E(\varphi)
	\Big),
\end{equation} 
wherein we have added the field potential $V(\varphi)$ in the Jordan frame, that transformed into  the potential $V_E$  in the Einstein frame,
$$
	V_E(\varphi)=\frac{V(\varphi)}{\Omega^2(\varphi)}.
$$
This potential accounts the non-minimal interaction, while the kinetic term is also deviates from the canonical form. The conformal transform reveals features of non-minimal interaction that is especially spectacular in theories with two scales: the Planckean scale and the scale of interaction as we will clearly see in the further study. 
\begin{figure}[h]
\begin{center}
\includegraphics[width=4.25in]{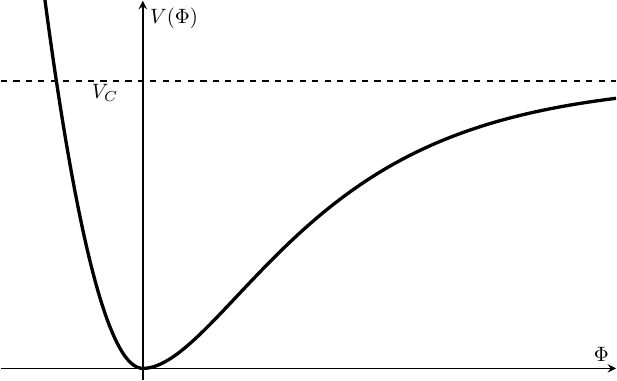}
\caption{Potentials for the free field and the inflation after the transition from the Jordan frame to the Einstein frame The dot with the arrow denotes the primary position of field in the non-stationary coherent state possessing a velocity $\dot\varphi>0$ at the bottom of potential.}
\label{jordan}
\end{center}
\end{figure}

Consider the case when $\tilde m_\mathrm{Pl}\gg \Lambda_\mathrm{int}$ and fields in the region $0< \varphi \ll \tilde m_\mathrm{Pl}^2/\Lambda_\mathrm{int}$ that implies
$$
	\frac{1}{\Omega(\varphi)}\ll \frac32\,\tilde m_\mathrm{Pl}^2
	\Bigg(\frac{\partial \ln\Omega}{\partial\varphi}\Bigg)^2.
$$
Therefore, in this region the non-canonical kinetic term equal to
$$
	\frac12(\partial_\mu\varphi)^2\cdot \frac32\,\tilde m_\mathrm{Pl}^2
	\Bigg(\frac{\partial \ln\Omega}{\partial\varphi}\Bigg)^2
$$
gets the canonical form under the field substitution
\begin{equation}
\label{int-7}
	\Phi=\tilde m_\mathrm{Pl}\sqrt{3/2}\ln\Omega(\varphi)=
	\tilde m_\mathrm{Pl}\sqrt{3/2}\ln
	\Big(1+\frac{\varphi\hspace*{2mm}}{\Lambda_\mathrm{int}}\Big).
\end{equation}
At the potential of free scalar field in the Jordan frame the potential of $\Phi$ equals 
\begin{equation}
\label{int-8}
	V_E=\frac12\,m^2\Lambda_\mathrm{int}^2\Bigg(1-
	\exp\bigg(-\frac{\Phi}{\tilde m_\mathrm{Pl}}\cdot\sqrt{2/3}\bigg)
	\Bigg)^2.
\end{equation}
It has the plateau 
\begin{equation}
\label{int-9}
	V_C=\frac12\,m^2\Lambda_\mathrm{int}^2,
\end{equation}
so that the field $\Phi$ is well suitable for the role of inflaton in the expansion of early universe \cite{Kallosh:2013tua}. The corresponding potential is shown in Fig. \ref{jordan}.

The square of inflaton mass at the bottom of potential equals 
\begin{equation}
\label{int-10}
	m_\mathrm{inf}^2=\frac23\,m^2\cdot\frac{\Lambda_\mathrm{int}^2}{\tilde m_\mathrm{Pl}^2}.
\end{equation}

We assume that the interaction with gravity conserves the vacuum contribution 
to the stress-energy tensor (\ref{cov-7}). However, this interaction disturbs the coherent state of scalar field in the way that the field at $\langle n \rangle\gg 1$ obeys the classical equation of motion with the evolution under the time $t$ instead of invariant $\tau$. Therefore, the model gives the suppression of vacuum density of energy and supports the inflation mechanism in the early universe. The evolution could start in the bottom of potential with a large velocity corresponding to the huge excitation of coherent state with $\langle n \rangle\gg 1$ as it shown in Fig. \ref{jordan}, while the interaction with gravity produces a friction of motion along the potential growth with further stopping and falling down under slow roll inflation from the plateau.

\section{Numerical estimates} \label{sec:NumEst}
Supposing the relevance of model parameters to the empirical values of observational cosmology we put  the vacuum density of energy and inflation plateau by the order of magnitude equal to 
$$
	\rho_\mathrm{vac}\sim (10^{-3}\mbox{ eV})^4,\qquad  V_C\sim (10^{16}\mbox{ GeV})^4, 
$$
while the one loop estimates with gravitons support the estimates
$$
	m\sim \Lambda\sim \Lambda_\mathrm{int}\sim V_C^{1/4}\sim 10^{16}\mbox{ GeV} \quad \Rightarrow \quad m_\mathrm{inf} \sim 10^{14}\mbox{ GeV}. 
$$
Then
\begin{equation}
\label{ave-1}
	\langle n\rangle\sim \frac{\tilde m_\mathrm{Pl}}{m} \sim \frac{\tilde m_\mathrm{Pl}}{\Lambda} \sim \frac{\Lambda}{m_\mathrm{inf}} \sim 250.
\end{equation}
This estimate gives the energy of initial coherent state in the reference 3D volume
\begin{equation}
\label{ave-2}
	\langle E(\varphi)\rangle\sim \langle n\rangle\, m \sim \tilde m_\mathrm{Pl}.
\end{equation}
So, the model deals with two energy scales naturally involved in the concordance model of universe inflation: the reduced Planckean mass and the scale of inflation plateausetting the cut-off and mass parameters.

\section{Conclusion}
We have shown that in the covariant formalism the bare value of zero-point energy density caused by the sub-Planckean cut-off scale can be suppressed under the following conditions: 
\begin{itemize}
\item the non-stationary coherent state of scalar field in the expanding universe,
\item the finite volume determined by the energy cut-off scale for the scalar field,
\item the coherent spatial-temporal structure of fluctuations in the invariant vacuum,
\item the natural estimates for the cut-off scale in comparison to the Planckean energy of scalar field in the finite volume.
\end{itemize}
The model involves at least two scales of energy: 
\begin{enumerate}
\item[(i)] the Planck scale of gravity, that also determines the energy of scalar field in the finite volume;
\item[(ii)] the sub-Planck cut-off setting the scale of finite volume as well as the scalar field mass, the field fluctuations in the vacuum and the plateau height of potential in models of inflationary early universe expansion relevant to the high precision data in the cosmology. 
\end{enumerate}
The relations between the scales as it is coming empirically require further studies. 

Note that other contributions to the vacuum energy density would be also suppressed due to the very non-stationary state of the scalar field since the additional terms $\rho_\mathrm{add}$ do contribute under averaging over the common vacuum state $|0_\mathrm{add}\rangle\otimes|0\rangle\,\mathrm e^{-\langle n\rangle/2}$. Therefore, other sources of suppressed vacuum energy density are significant only if  $\rho_\mathrm{add}\sim (10^{16}\mbox{ GeV})^4 $. In this case the sign of actual cosmological constant depends on resulting sum over all relevant terms.

\bibliography{bib-CC4D}
	
\end{document}